\documentclass{article}
\usepackage{amsmath,amsthm,amsfonts,amscd,amssymb} 
\usepackage[hmargin=2cm, vmargin=3cm]{geometry}
\usepackage{graphicx}
\usepackage{mathrsfs}
\usepackage{titlesec}
\usepackage{authblk}

\usepackage{bm}
\usepackage{graphicx}
\usepackage{indentfirst}
\usepackage{color}
\usepackage{algpseudocode}
\usepackage{enumerate}

\usepackage{hyperref}
\hypersetup{
    colorlinks=true,
    linkcolor=blue,
    filecolor=magenta,      
    urlcolor=cyan,
    pdftitle={Overleaf Example},
    pdfpagemode=FullScreen,
    }

\usepackage{amssymb}
\usepackage{amsmath}

\newtheorem*{remark}{Remark}

\def\R{\mathbb{R}}

\def\R{\mathbb{R}}
\def\erf{{\rm erf}}

\DeclareMathAlphabet{\mathscr}{U}{dutchcal}{m}{n}
\SetMathAlphabet{\mathscr}{bold}{U}{dutchcal}{b}{n}
\DeclareMathAlphabet{\mathbscr} {U}{dutchcal}{b}{n}

\graphicspath{{./Figures1/}}
\begin{document}

\title{\bf{Candidate Voter Dynamicsl}}

\author[1]{Christoph Borgers \thanks{christoph.borgers@tufts.edu}}
\author[1]{Natasa Dragovic \thanks{natasa.dragovic@stthomas.edu}}
\author[3]{Arkadz Kirshtein \thanks{arkadz.kirshtein@tamucc.edu}}
\affil[1]{Tufts University}
\affil[2]{University of Saint Thomas}
\affil[3]{Texas A\& M Corpus Christi}



\date{19 November 2025}
\maketitle

\begin{abstract}
We model dynamically changing candidate positions in the face of a dynamic electorate. 
To formulate our equations, we use  a space-time-continuous Hegselmann-Krause equation, 
which we solve using a particle method. We use the combined candidate-voter model
to demonstrate the possibility of discontinuous jumps in candidate behavior as parameters
of the model are varied. We also extend the analysis to a three candidate scenario. We observe that depending on the parameters, candidates do not always come or stay together at their dynamically evolving position.

\end{abstract}

\section{Introduction.} 
People's opinions and beliefs are influenced in complex ways by families, friends, colleagues, media, as well
as politicians and other mega-influencers \cite{Acemoglu_opinion,Bisin_2000,Bisin_2001,haensch2023,Boyd_1985,Cavalli_1981}.
In recent decades, attempts have been made to understand aspects of this process using mathematical modeling and computational simulation; for surveys on opinion dynamics, see for instance
  \cite{Aydogdu_2017,Anderson_2019,Lorenz_2007,Mossel_2017,Proskurnikov_2017,Proskurnikov_2018}.

Many models of opinion dynamics are based on the assumption  that we are influenced more easily by people
whom we {\em almost} agree with already than by those whose views starkly differ from ours.
A similar but more general phenomenon is known as {\em biased assimilation} among psychologists --- our tendency 
to filter and interpret information in such a way that it supports our preconceived notions \cite{Lord_1979}. Models of opinion dynamics
 based on this assumption are known as {\em bounded confidence models} \cite{Aydogdu_2017,Canuto_2012,Mirtabatabaei_2012}. 
A popular example  is due to Hegselmann and Krause \cite{hegsel_krause_2002,Hegselmann_Krause_2005}, 
building on earlier work by Krause \cite{krause_1997,krause_2000}. It has been studied extensively
in the literature (see
for instance \cite{Lorenz_2005, Lorenz_2006, Blondel_2009,Perrier_2024}), and will be our starting point
here. 

The original Hegselmann-Krause model is discrete
in both time and opinion space: A finite number of agents move one step
at a time. Similar models that are
continuous in time \cite{Piccoli_2021}, 
opinion space \cite{Wedin_2015}, or opinion space and time \cite{Borgers_et_al_conference_paper,Goddard_2022} have been proposed since. We focus
here on a fully continuous model. This allows us to add the dynamics of candidates in a natural optimization-motivated way, with the 
candidates following ordinary differential equations; see equations (\ref{eq:delldt}) below. 

In recent years, there have been several studies of how opinion dynamics interact with the dynamics of political candidates and with voter behavior\cite{Cahill_2024, tomlinson2024replicating,Jones_et_al_on_abstention}. The work presented here contributes to this new field of study, merging ideas from three earlier papers \cite{Borgers_et_al_candidate_dynamics,Borgers_codee, Borgers_et_al_conference_paper}. In \cite{Borgers_et_al_candidate_dynamics, Borgers_codee}, papers studied
the response of dynamic candidates
to a  static electorate. The main result was that 
the optimal strategy for a candidate can, under certain conditions specified in \cite{Borgers_et_al_candidate_dynamics}, depend
discontinuously on the voters' ``loyalty", their willingness to not abstain and vote for the ``lesser of two evils". Larger $\gamma$ values
prompt voters to consider candidates further from themselves in the opinion
space. While lower gamma values make voter more close-minded and likely to abstain. In \cite{Borgers_et_al_conference_paper}, they developed a fully continuous model
of the dynamics of the electorate, in the absence of political candidates,  and a particle method for solving the
equation. Here, we describe a more complex model in which both the voters and the candidates are dynamic. 

The continuous model used here is a variation of \cite{Borgers_et_al_conference_paper}. For this reason, and also for
the reader's convenience, we briefly repeat the derivation of the model here. Adding the candidate dynamics model from
\cite{Borgers_et_al_candidate_dynamics}, we again find discontinuous dependence on voter loyalty, but now also on
other model parameters.

We speculate that the turbulent recent U.S.\ presidential politics may reflect discontinuities similar to those described here. Perhaps many Americans became susceptible to the
suggestion that only fraud could explain the 2020 results because winning strategies shifted dramatically from
one election cycle to the next.

The paper is structured as follows. We start with the particle model of voter opinion evolution. We then add interaction with first two, then three candidates. We conclude with an outline of possible further extensions.

\section{Voter dynamics.} 
\label{sec:particle_model} We first describe a particle model of opinion dynamics, then interpret it as a discretization of  a fully continuous model. 
 We assume that any individual's opinions can be characterized by a single real number $x$. 
Let $X_1,X_2,\ldots,X_n \in \R$, and assume for now that the $X_i$ are the only opinions represented in the electorate. 
If $w_i$ is the fraction of individuals who hold opinion $X_i$, then the ``density" of opinions altogether is the distribution
\begin{equation}
\label{eq:weighted_sum_of_diracs}
 \sum_{i=1}^n w_i \delta(x-X_i),
\end{equation}
where $\delta$ denotes the Dirac delta distribution. The condition that this be a probability measure becomes
$
\sum_{i=1}^n w_i =1.
$

Any Borel probability measure $\mu$ on the real line can  be approximated arbitrarily well, in the distributional sense,
by a weighted sum of delta functions in the form (\ref{eq:weighted_sum_of_diracs}). In fact, let $n \geq 1$ be an integer,
$\Delta x >0$ a real number, and
define, for all integers $i$ with $-n+1 \leq  i  \leq n-1$, 
\begin{equation}
\label{weights}
w_i = \frac{\mu \left[ \left( i - \frac{1}{2} \right) \Delta x, \left( i + \frac{1}{2} \right) \Delta x \right) }{\displaystyle{\sum_{k=-n+1}^{n-1} } \mu \left[ \left( k - \frac{1}{2} \right) \Delta x, \left( k + \frac{1}{2} \right) \Delta x \right) }.
\end{equation}
Then 
\begin{equation}
\label{dirac_sum}
\sum_{i=-n+1}^{n-1}  w_i   \delta \left( x - i \Delta x \right)
\end{equation}
converges weakly to $\mu$ if $n \rightarrow \infty$ and $\Delta x \rightarrow 0$ in such a way 
that $n \Delta x \rightarrow \infty$.
We assume that the opinion holders in the $i$-th cluster, that is, opinion holders
with opinion $X_i$, find the opinions represented by other clusters persuasive especially if the other clusters are large and/or nearby in opinion
space:
\begin{equation}
\label{time_evolution}
\frac{dX_i}{dt} = 
\sum_{j=1}^n \eta(|X_j-X_i|) w_j (X_j-X_i)
\end{equation}
where 
$$
\eta: ~~  [0,\infty) \rightarrow [0,1]
$$
is a decreasing function with $\lim_{z \rightarrow \infty} \eta(z) = 0$, called the 
{\em interaction function}. 
We use
\begin{equation}
\label{eq:def_eta}
\eta(z) = \max \left( 1 - \frac{z}{\nu}, 0 \right) ~~~ \mbox{for $z \geq 0$}
\end{equation}
where $\nu>0$ is a parameter determining how broad-minded the opinion holders are. Larger $\nu$ means greater broad-mindedness. 
\begin{remark}
The introduction of $\eta$ reflects the assumption that the opinion of agents is influenced more by those closer to the opinion space, while it can still be influenced by differing opinions but to a lesser degree. There are many similar models in the literature, differing in details only; see for instance \cite{Aydogdu_2017,Blondel_2009,Canuto_2012}.
\end{remark}
 
We always use the explicit midpoint method with a fixed time step $\Delta t$ to solve Eq.\  (\ref{time_evolution}). 
We obtain at each time step 
a weighted sum of delta functions in the form (\ref{eq:weighted_sum_of_diracs}), which can in turn be approximated 
arbitrarily well by 
a smooth density. For instance, the smooth probability density
\begin{equation}
\label{smooth}
\sum_{i=1}^n w_i  \frac{e^{- (x-X_i)^2/(2 \sigma^2)}}{\sqrt{ 2 \pi \sigma^2}} 
\end{equation}
converges to (\ref{eq:weighted_sum_of_diracs}), in the distributional sense, as $\sigma \rightarrow 0$. 
We view $\sigma>0$ as another numerical parameter.

\begin{figure}[h!] 
\begin{center}
\includegraphics[width=0.6\textwidth]{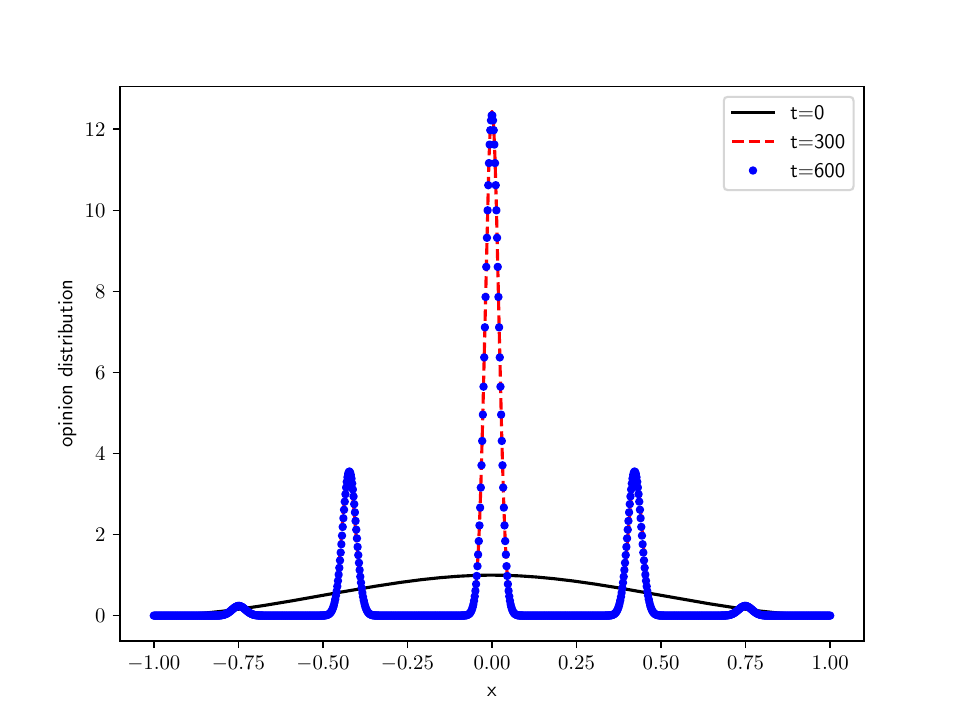}
\caption{Black: opinion density at the initial time. Red and blue: opinion density at time $T=300$ with $n=100$ and $\Delta t = 2$ (red) and 
at time $T=600$ with $n=200$ and $\Delta t=1$ (blue). Other parameters were $\nu=0.25$ and $\sigma=0.02$.}
\label{fig:CONSERVATIVE_HK}
\end{center}
\end{figure}

As an example,  we construct a compactly supported differentiable probability density on the interval $[-1,1]$ as follows. 
First, let $p(x)$  be the uniquely determined cubic polynomial with $p(0)=1$, $p'(0)=p(1)=p'(1)=0$: 
$$
p(x) = 2 x^3 - 3x^2+1.
$$
Define $f(x) = p(x)$ for $x \in [0,1]$,  $f(x) = p(-x)$ for $x \in [-1,0]$, and $f(x) = 0$ outside of $[-1,1]$, or briefly:
\begin{equation}
\label{eq:initial_density}
f(x) =  \left\{ \begin{array}{cl} 2 x^2|x| - 3 x^2+1 & \mbox{if $|x| \leq 1$}, \\
0 & \mbox{otherwise};
\end{array}
\right.
\end{equation}
see the black curve in Fig.\ \ref{fig:CONSERVATIVE_HK}. Let
$$
X_i(0) = -1 + \left(i - \frac{1}{2} \right) \Delta x, ~~~i =1,\ldots,n, ~~~\Delta x = \frac{2}{n}. 
$$
The value of $n$ will be specified shortly. In general, as noted following eq.\ (3), $n \Delta x$ must tend to infinity for convergence. If this were not the case, the tail of the initial distribution would never be taken into consideration. Here, however, the initial distribution is compactly supported. This is why here it is not necessary for $n \Delta x$ to tend to infinity. 

Let the 
$w_i$  be proportional to $f(X_i(0))$, 
with the constant of proportionality chosen to make the sum of the $w_i$ equal to $1$.
We solve the equations using the midpoint method with a time step $\Delta t$ up to a time $T$ (both $\Delta t$ and $T$ will be specified shortly), then plot
the function
$$
f(x,T) = 
\sum_{i=1}^{n} w_i  \frac{e^{-(x-X_i(T))^2/(2 \sigma^2)}}{\sqrt{2 \pi \sigma^2}}.
$$

Figure  \ref{fig:CONSERVATIVE_HK} shows results obtained with $n=100$, $\Delta t = 2$, $T=300$ (red) and 
then again with $n=200$, $\Delta t =1$, $T=600$ (blue). The red and blue results are indistinguishable to the eye --- so 
the calculations are likely to yield the ``fully resolved" final state, in the sense that higher resolution, or larger $T$, would not result in visibly different pictures.
In this calculation, $\nu=0.25$. The value of $\nu$ of course affects the outcome, for instance for $\nu=0.75$, global consensus is reached. Also, $\sigma=0.02$ in Figure \ref{fig:CONSERVATIVE_HK} was taken to match with the larger value of $\Delta x$ from the initial distributions in the simulations. Smaller $\sigma$ would result in tighter peaks, but this is merely a matter of graphics; the red
and blue curves in Figure 
\ref{fig:CONSERVATIVE_HK} should be interpreted as showing a sum of five weighted delta distributions.

Our particle model can be viewed as a discretization of the following equation for a continuous opinion density $f(x,t)$: 
\begin{equation}
\label{eq:differential_integral_equation}
f_t(x,t) + \left(c(x,t)  ~\! f(x,t) \right)_x=0 
\end{equation}
with 
\begin{equation}
\label{eq:velocity}
c(x,t) = \int_{-\infty}^{\infty} \eta(|y-x|) f(y,t) (y-x) dy. 
\end{equation}
Equation (\ref{eq:differential_integral_equation}) describes, in general, the motion of mass in a velocity field $c(x,t)$; see any introductory book on partial
differential equations. Notice the analogy 
between Eq.\ (\ref{eq:velocity}) and the right-hand side of (\ref{time_evolution}). The term $\eta(|X_j-X_i|)$ is analogous to $\eta(|y-x|)$, $w_j$ corresponds
to $f(y,t) dy$, and $X_j-X_i$ corresponds to $y-x$. Summation in (\ref{time_evolution}) becomes integration in (\ref{eq:velocity}). Our particle method
approximates the convolution $f \ast \varphi_\sigma$, where $\varphi_\sigma$ denotes the Gaussian density with mean $0$ and variance $\sigma^2$.
For an analogous particle discretization of a very similar fully continuous equation, weak second-order convergence in both time and opinion space was verified in \cite{Borgers_et_al_conference_paper}. We will not repeat that here.

\section{Candidate-voter dynamics.} Consider two candidates competing
for votes while opinions in the electorate are changing. 
Our modeling here follows \cite{Borgers_et_al_candidate_dynamics}, again with slight simplifications.
We consider two political candidates, $L$ and $R$, with political positions $\ell$ and $r$, where $\ell<r$. We will 
study how the candidates might shift positions to increase their chance of winning elections. 
However, in this model we assume voter dynamics to be independent of the candidate behavior.

We assume that a voter whose position $x$ is closer to $\ell$ than to $r$ will vote for $L$, unless they abstain. Their probability
of voting (not abstaining) is taken to be 
\begin{equation}
\label{eq:def_gamma}
e^{-(x-\ell)^2/(2 \gamma^2)}
\end{equation}
 where $\gamma>0$ is a new model parameter measuring voter loyalty. Higher $\gamma$ results in higher probability of voting for a far-away candidate. Consequently the fraction of voters who 
will vote for $L$, if the election is held at time $t$, will be 
\begin{equation}
\label{eq:S_L}
S_L = S_L(\ell,r,t) = \int_{-\infty}^{(\ell+r)/2} f(x,t) e^{-(x-\ell)^2/(2 \gamma^2)} dx, 
\end{equation}
where $f$ denotes the probability density of voter positions. Analogously, 
\begin{equation}
\label{eq:S_R}
S_R = S_R(\ell,r,t) = \int_{(\ell+r)/2}^\infty f(x,t) e^{-(x-r)^2/(2\gamma^2)} dx.
\end{equation}
We assume that $L$ and $R$ optimize their position using continuous steepest descent: 
\begin{equation}
\label{eq:delldt}
\frac{d \ell}{dt} = \alpha \frac{\partial S_L}{\partial \ell}(\ell, r, t), ~~~ \frac{dr}{dt} = \beta \frac{\partial S_R}{\partial r} (\ell, r, t), 
\end{equation}
where $\alpha$ and $\beta$ are parameters measuring the
eagerness with which $L$ changes positions opportunistically. 

 The voter distributions computed by
our method are of the form 
$$
f(x,t) = \sum_{i=1}^n w_i  \frac{e^{- (x-X_i(t))^2/(2 \sigma^2)}}{\sqrt{ 2 \pi \sigma^2}}.
$$
(Compare Eq.\ 
(\ref{smooth}).) Using this formula, we can explicitly evaluate $S_L$, $S_R$, and the requisite partial derivatives.
With 
\begin{equation}
\label{eq:def_xi}
\erf(z) = \frac{2}{\sqrt{\pi}} \int_0^z e^{-s^2} ds
~~~~~~\mbox{and} ~~~~~~~  \xi = \sqrt{ \gamma^2 +  \sigma^2}~\!, 
\end{equation} 
we obtain
{\scriptsize
\begin{eqnarray}
\label{eq:S_L_evaluated} 
S_L(\ell,r,t) &=&  \gamma  \sum_{i=1}^n  w_i ~\! \frac{  e^{- \frac{(X_i-\ell)^2}{2\xi^2}}}{ 2 \xi} \left[
1 - \erf \left( - \frac{   \frac{\gamma}{\sigma} \left( \frac{\ell+r}{2}  -  X_i \right) + \frac{\sigma}{\gamma}  \frac{r-\ell}{2} }{\sqrt{2}~\!  \xi} \right)
\right], \\
\label{eq:S_R_evaluated}
S_R(\ell,r,t) &=&
\gamma  \sum_{i=1}^n  w_i ~\! \frac{  e^{- \frac{(X_i-r)^2}{2 \xi^2}}}{ 2\xi} \left[
1 + \erf \left(  - \frac{   \frac{ \gamma}{\sigma} \left(  \frac{\ell+r}{2} -X_i \right) - \frac{\sigma}{ \gamma} \frac{r-\ell}{2}  }{ \sqrt{2} ~\! \xi} \right)
\right], \\
\nonumber
\frac{\partial S_L}{\partial \ell}(\ell,r,t) &=& 
\gamma  \sum_{i=1}^n  w_i ~\! \frac{  e^{- \frac{(X_i-\ell)^2}{2 \xi^2}}(X_i-\ell)}{2 \xi^3}
~\!  \left[
1 - \erf \left( - \frac{   \frac{\gamma}{\sigma} \left( \frac{\ell+r}{2}  -  X_i \right) + \frac{\sigma}{ \gamma} \frac{r-\ell}{2} }{\sqrt{2} ~\! \xi} \right)
\right]\\
&&+ \gamma  \sum_{i=1}^n  w_i ~\! \frac{  e^{- \frac{(X_i-\ell)^2}{2 \xi^2}}}{ \sqrt{8\pi} ~\! \xi^2}  ~
\left(   \frac{\gamma}{ \sigma} - \frac{\sigma}{ \gamma} \right) 
 \exp \left[ - \frac{ \left(  \frac{ \gamma}{\sigma} \left( \frac{\ell+r}{2}  -  X_i \right) + \frac{\sigma}{ \gamma}  \frac{r-\ell}{2}  \right)^2}{ 2 ~\! \xi^2} \right], \label{eq:dS_L_evaluated}\\
\nonumber
\frac{\partial S_R}{\partial r}(\ell,r,t) &=& 
 \gamma  \sum_{i=1}^n  w_i ~\! \frac{  e^{- \frac{(X_i-r)^2}{2\xi^2}}(X_i-r)}{ 2\xi^3}
~\!  \left[
1 + \erf \left(  -\frac{   \frac{ \gamma}{\sigma} \left(  \frac{\ell+r}{2} -X_i \right) - \frac{\sigma}{\gamma} \frac{r-\ell}{2}}{ \sqrt{2} ~\! \xi} \right)
\right]\\
&&- \gamma  \sum_{i=1}^n  w_i ~\! \frac{  e^{- \frac{(X_i-r)^2}{2\xi^2}}}{ \sqrt{8\pi} ~\! \xi^2}  ~
\left(  \frac{\gamma}{ \sigma} - \frac{\sigma}{\gamma} \right) 
 \exp \left[ - \frac{ \left(  \frac{\gamma}{\sigma} \left(  \frac{\ell+r}{2} - X_i  \right) - \frac{\sigma}{\gamma} \frac{r-\ell}{2} \right)^2}{ 2 \xi^2} \right].\label{eq:dS_R_evaluated}
\end{eqnarray}}

\subsection{Dynamics after collision}

In reference \cite{Borgers_et_al_candidate_dynamics}, the simulation was stopped if the two candidate positions came together. However, in our setting, the electorate changes with time as well, and it therefore seems sensible to allow the candidate positions to split apart again at a later time. We do not, however, allow the two candidates to move past each other. Thus we set:
\begin{align}
    \frac{d \ell}{dt} = \frac{dr}{dt} =0.5\left( \alpha \frac{\partial S_L}{\partial \ell}(\ell, r, t)+ \beta \frac{\partial S_R}{\partial r} (\ell, r, t)\right), \quad \qquad
\nonumber\\ \text{while}  \quad \alpha \frac{\partial S_L}{\partial \ell}(\ell, r, t)> \beta \frac{\partial S_R}{\partial r} (\ell, r, t).
\end{align}
As soon as $\alpha \frac{\partial S_L}{\partial \ell} < \beta \frac{\partial S_R}{\partial r}$, the two candidate positions split apart and are again governed by the differential equations in (13).

\begin{remark}
    There are several alternative strategies for handling candidate collisions, including but not limited to letting them pass each other or making them stationary until their collision is resolved. Each of these strategies has its own implications for the dynamical system, particularly in how they affect the stability and the trajectories of other agents in the vicinity. Future work may explore and compare the merits of these approaches through simulation. However, we currently settled on a method that allows for a more intuitive handling of the vote shares during and after collision.
\end{remark}

Once their preferences diverge, their positions can split apart, and they can resume their independent trajectories, each adjusting their velocities according to their individual desired rates. This approach allows candidates to respond dynamically to the shifting preferences of the electorate, facilitating a more adaptive political strategy. It should be noted that the reason for their preferences to diverge could stem from voter openness that affects optimal positions as the electorate continues to evolve.

\section{Simulations} 

The model has a fairly large number of parameters, and at present we have no
overall picture of the dependence of candidate behavior on these parameters.
Here we merely give examples of discontinuous dependence 
on the parameters:
voter open-mindedness $\nu$ (eq. \eqref{eq:def_eta}), voter loyalty $\gamma$ (eq.\ \eqref{eq:def_gamma}),  and degree of opportunism of the candidates $\alpha$ and $\beta$ (eq.\ \eqref{eq:delldt}).


What can behave discontinuously on these parameters is the {\em eventual outcome} of the process of candidate position optimization. The solutions of the differential equations
at a fixed time must, of course, by a general theorem about ordinary differential equations depend continuously on parameters.

The possibility of discontinuous dependence of the outcome of opportunistic position optimization on voter loyalty had already been
pointed out in \cite{Borgers_et_al_candidate_dynamics}, but there the opinion distribution in the electorate was taken to be 
fixed. In \cite{Borgers_et_al_candidate_dynamics}, discontinuous behavior was only observed for a fixed bimodal electorate. 
Here we always let the initial opinion density be  the function given in (\ref{eq:initial_density}) and indicated in black in Fig.\ \ref{fig:CONSERVATIVE_HK}. Multiple peaks emerge as a result of opinion dynamics, as shown in Fig.\ \ref{fig:CONSERVATIVE_HK}.

\begin{figure}[h!] 
\begin{center}
\includegraphics[width=0.6\textwidth]{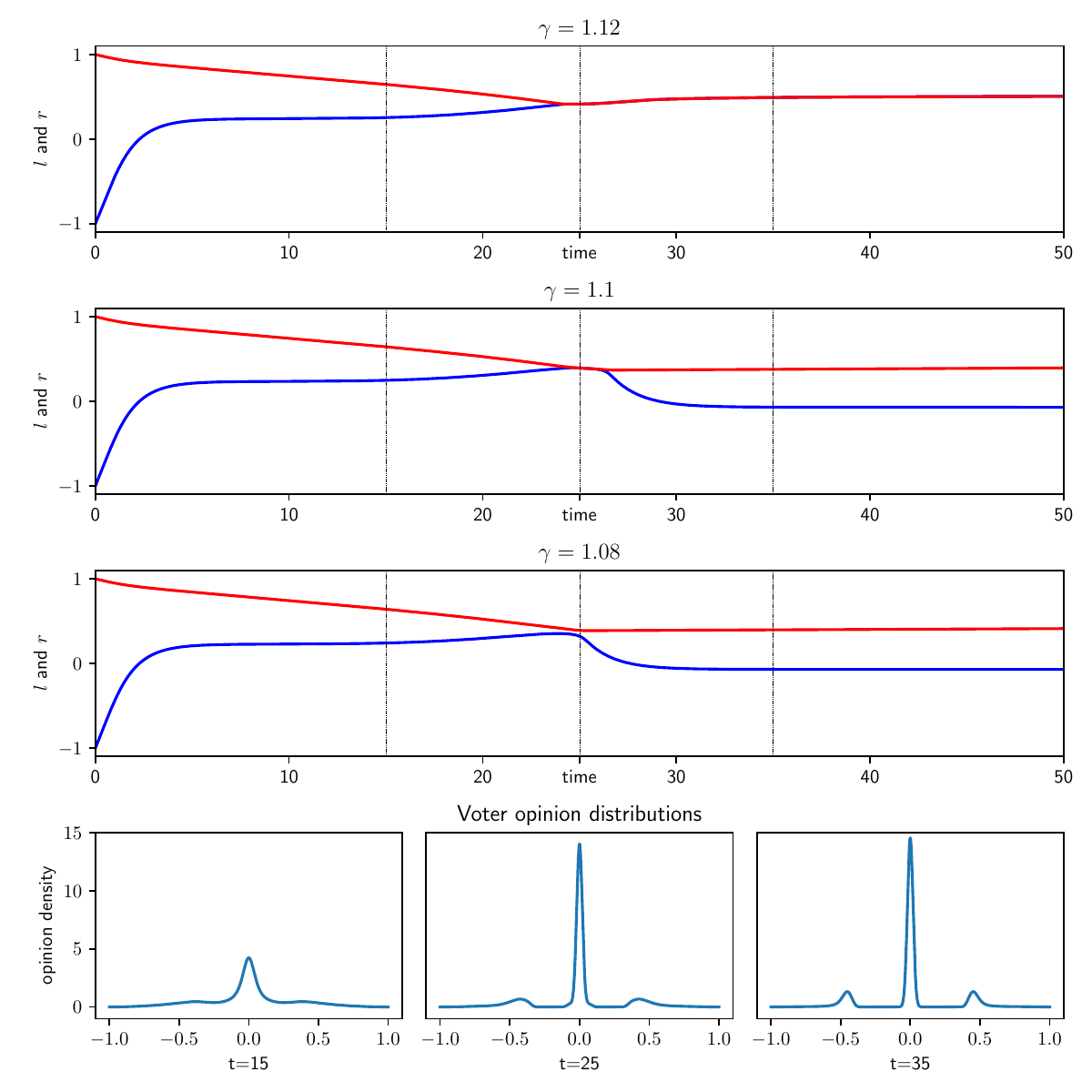}
\caption{\textbf{Varying voter loyalty.} Positions taken by left and right candidates as a function of time, with $\nu=0.30$, $\alpha=1$, $\beta=0.1$, and three different values of $\gamma$. Horizontal axis represents time. Vertical lines at $t=15$, $t=25$, and $t=35$ in the three upper panels indicate the moments at which the state of the electorate is respectively shown on the lowest panels. } 
\label{fig:VOTER_LOYALTY}
\end{center}
\end{figure}

In Figure~\ref{fig:VOTER_LOYALTY}, we fixed voter open-mindedness $\nu = 0.30$, candidate opportunism $\alpha=1$ and $\beta=0.1$, and varied voter loyalty $\gamma$. For sufficiently large $\gamma$, the two candidates coalesce rapidly.
 For smaller $\gamma$, $\ell$ and $r$ approach each other, but eventually, due to increasing polarization
of the electorate, they turn away from
each other. 
The lower panel of the Figure~\ref{fig:VOTER_LOYALTY} shows the electorate at $t=15$, $t=25$, and $t=35$ respectively from the left to the right. Recall that in our model the opinion dynamics in the electorate does not respond to the candidates.
\begin{figure}[h!] 
\begin{center}
\includegraphics[width=0.6\textwidth]{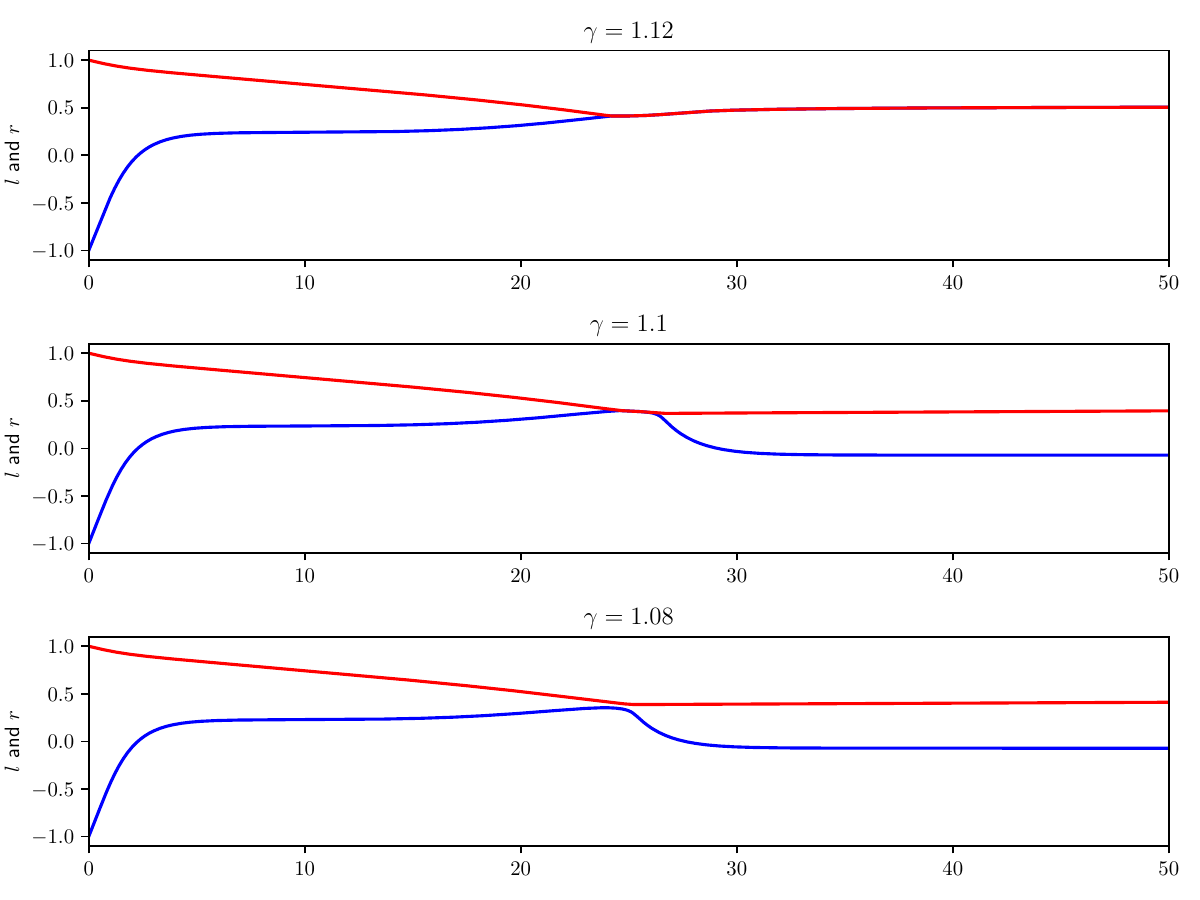}

\caption{\textbf{Final candidate positions for varying voter loyalty.} Positions taken by left and right candidates at time $t=50$ as a function of voter loyalty $\gamma$, with $\nu=0.30$, $\alpha=1$, $\beta=0.1$. } 
\label{fig:FINAL_CANDIDATE_POSITION}
\end{center}
\end{figure}


In Figure\ \ref{fig:FINAL_CANDIDATE_POSITION} we plot the final candidate positions as a function of voter loyalty. We show the positions taken by left and right candidates at time $t=50$ as a function of $\gamma$, with $\nu=0.30$, $\alpha=1$, $\beta=0.1$. For $\gamma>1.12$ we observe the two candidates coming together. Notice that the dependence of the final position on voter loyalty is discontinuous.  

\begin{figure}[ht!] 
\begin{center}
\includegraphics[width=0.6\textwidth]{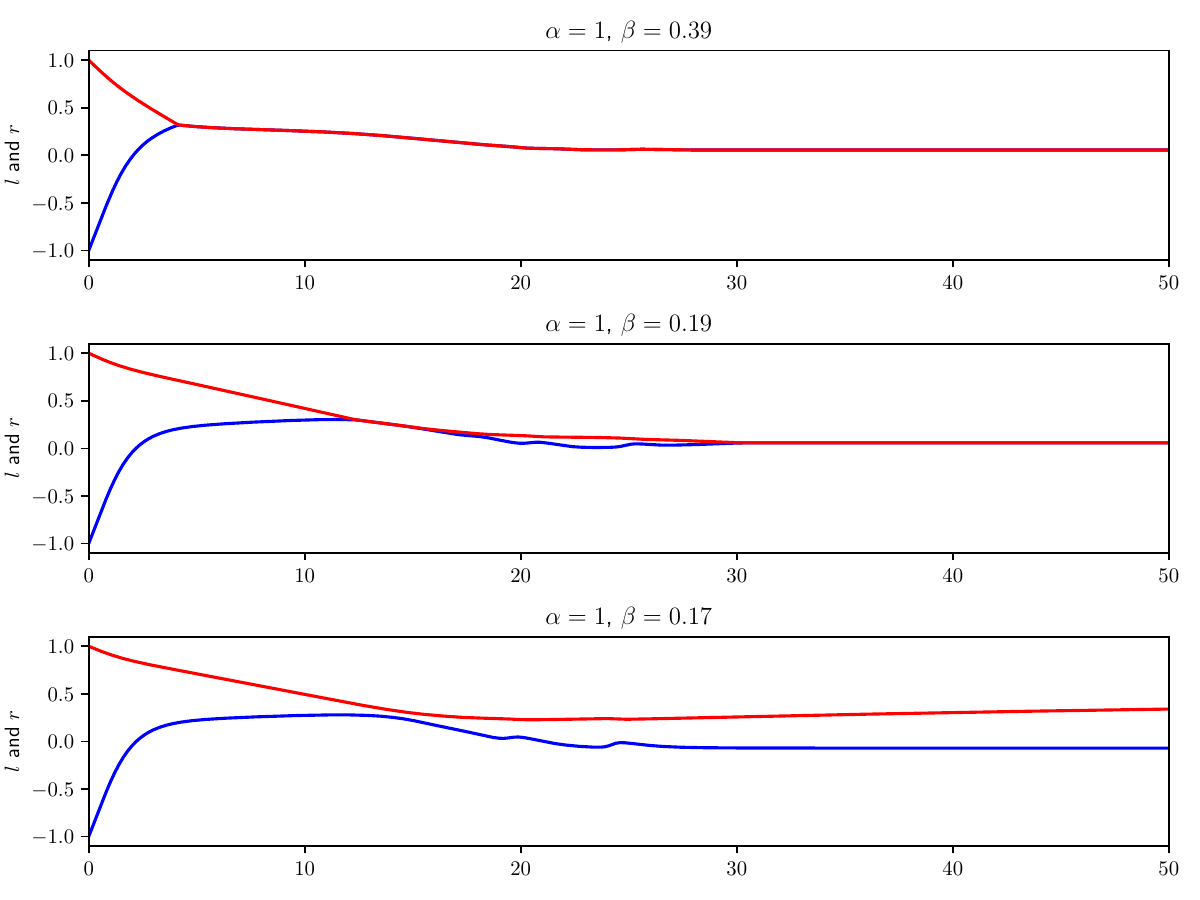}
\caption{\textbf{Varying voter open-mindedness.} Positions taken by left and right candidates as a function of time, with $\alpha=1$, $\beta=0.1$, $\gamma=1$, and three different values of $\nu$.} 
\label{fig:VOTER_OPENMINDEDNESS}
\end{center}
\end{figure}

In Figure\ \ref{fig:VOTER_OPENMINDEDNESS}, we fixed  candidate opportunism $\alpha=1$ and $\beta=0.1$, voter loyalty $\gamma=1$, and vary voter open-mindedness $\nu$. 
The result is a bit surprising: Greater $\nu$, that is, higher voter open-mindedness, causes the candidate positions to move apart from
each other, while smaller $\nu$, that is, lower voter open-mindedness, causes them to coalesce. The reason is that for the larger value of $\nu$, there will be slightly more 
centrist voters eventually. To capture the center without losing too many left-wing voters, the left-wing candidate moves back towards the center as the centrist camp grows more
pronounced. 

\begin{figure}[h!] 
\begin{center}
\includegraphics[width=0.6\textwidth]{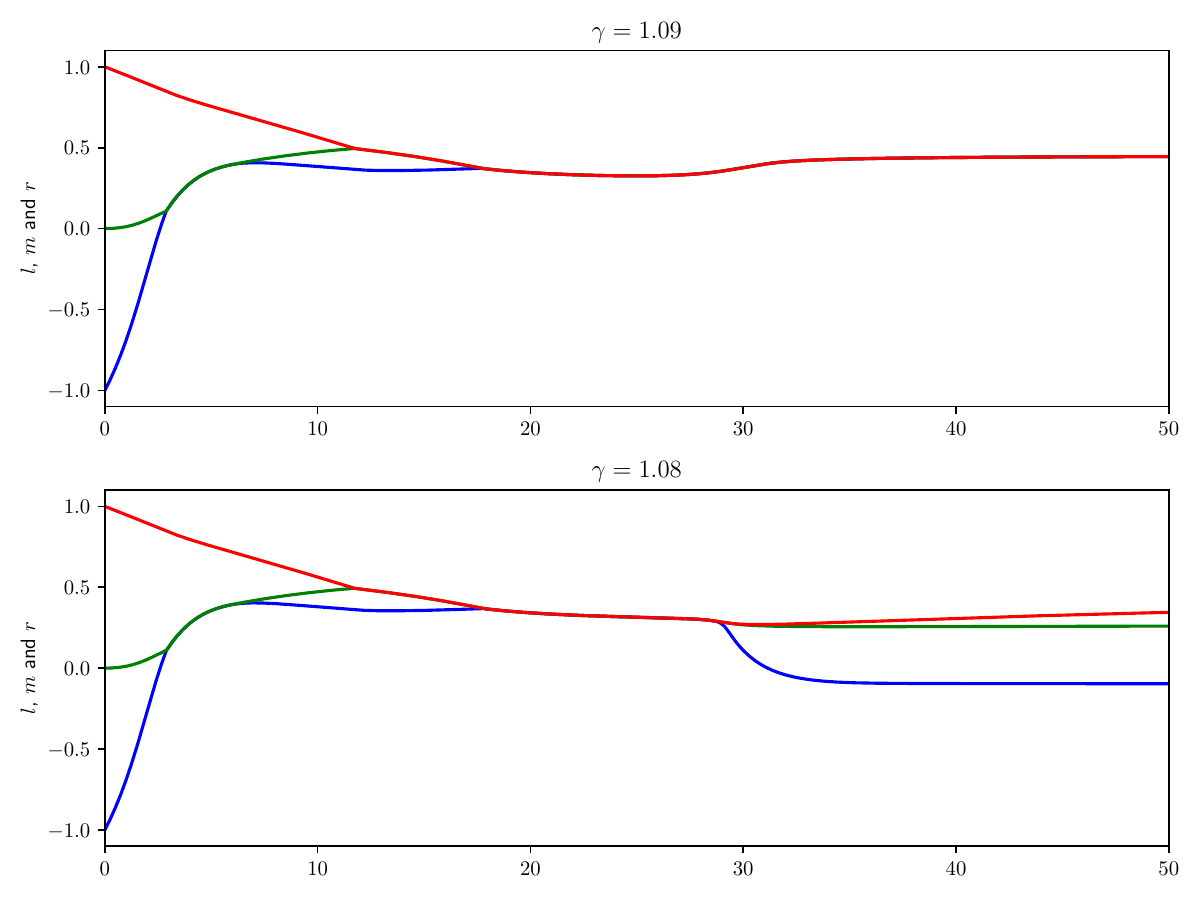}
\caption{\textbf{Varying candidate opportunism.} Positions taken by left and right candidates as a function of time, with $\nu=0.30$, $\gamma=1$, $\alpha=1$, and three different values of $\beta$.} 
\label{fig:CANDIDATE_OPPORTUNISM}
\end{center}
\end{figure}

In Figure~\ref{fig:CANDIDATE_OPPORTUNISM}, we fixed voter open-mindedness $\nu = 0.30$, voter loyalty $\gamma=1$, left candidate opportunism $\alpha=1$, and varied right candidate opportunism $\beta$.
For sufficiently large $\beta$, the two candidates coalesce rapidly. For small $\beta$, $\ell$ and $r$ approach each other, but eventually, due to increasing polarization
of the electorate, they turn away from
each other. 

\newpage

\section{Effect of a third candidate} 
The introduction of a third candidate into the electoral model alters the dynamics of candidate positioning significantly. In this section, we analyze the implications of this additional candidate on the strategies of the existing candidates and the resulting dynamics. The presence of a third candidate can lead to a fragmentation of the voter base, as candidates may attempt to differentiate themselves more sharply to capture distinct segments of the electorate. 

\subsection{Model with 3 candidates}
With the introduction of a third candidate we need to adjust the voter shares earned by each candidate. Let us assume the three political candidates L, M, and R with positions $\ell$, $m$, and $r$, where $\ell<m<r$. Then their corresponding voter shares will be 

\begin{eqnarray}
S_L = S_L(\ell,m,r,t) &=& \int_{-\infty}^{(\ell+m)/2} f(x,t) e^{-(x-\ell)^2/(2 \gamma^2)} dx, \\
S_M = S_M(\ell,m,r,t) &=& \int_{(\ell+m)/2}^{(m+r)/2} f(x,t) e^{-(x-m)^2/(2 \gamma^2)} dx, \\
S_R = S_R(\ell,m,r,t) &=& \int_{(m+r)/2}^{\infty} f(x,t) e^{-(x-r)^2/(2 \gamma^2)} dx.
\end{eqnarray}

Consequently, the equation for candidates' position optimizations become
{\small\begin{equation}
    \frac{d\ell}{dt}=\alpha\frac{\partial S_L}{d\ell}(\ell,m,r,t),\quad
    \frac{dm}{dt}=\kappa\frac{\partial S_M}{dm}(\ell,m,r,t),\quad 
\frac{dr}{dt}=\beta\frac{\partial S_R}{dr}(\ell,m,r,t).
\end{equation}}

Then using voter distribution $$
f(x,t) = \sum_{i=1}^n w_i  \frac{e^{- (x-X_i(t))^2/(2 \sigma^2)}}{\sqrt{ 2 \pi \sigma^2}}.
$$
with definitions \eqref{eq:def_xi} we obtain
{\footnotesize\begin{eqnarray}
\label{eq:3C-S_L_evaluated}
S_L(\ell,m, r,t) &=&  \gamma  \sum_{i=1}^n  w_i ~\! \frac{  e^{- \frac{(X_i-\ell)^2}{2\xi^2}}}{ 2 \xi} \left[
1 - \erf \left( - \frac{   \frac{\gamma}{\sigma} \left( \frac{\ell+m}{2}  -  X_i \right) + \frac{\sigma}{\gamma}  \frac{m-\ell}{2} }{\sqrt{2}~\!  \xi} \right)
\right], \\
 S_M(\ell,m, r,t) &=&  \gamma  \sum_{i=1}^n  w_i ~\! \frac{  e^{- \frac{(X_i-m)^2}{2\xi^2}}}{ 2 \xi} \left[
\erf \left( - \frac{   \frac{\gamma}{\sigma} \left( \frac{\ell+m}{2}  -  X_i \right) - \frac{\sigma}{\gamma}  \frac{m-\ell}{2} }{\sqrt{2}~\!  \xi} \right)\right.\nonumber\\
&&\qquad\qquad\qquad\qquad\quad\left.
- \erf \left( - \frac{   \frac{\gamma}{\sigma} \left( \frac{r+m}{2}  -  X_i \right) + \frac{\sigma}{\gamma}  \frac{r-m}{2} }{\sqrt{2}~\!  \xi} \right)
\right], \\
S_R(\ell,m,r,t) &=&
\gamma  \sum_{i=1}^n  w_i ~\! \frac{  e^{- \frac{(X_i-r)^2}{2 \xi^2}}}{ 2\xi} \left[
1 + \erf \left(  - \frac{   \frac{ \gamma}{\sigma} \left(  \frac{m+r}{2} -X_i \right) - \frac{\sigma}{ \gamma} \frac{r-m}{2}  }{ \sqrt{2} ~\! \xi} \right)
\right],
\end{eqnarray}}
{\scriptsize\begin{eqnarray}
\nonumber
\frac{\partial S_L}{\partial \ell}(\ell,m, r,t) &=& 
\gamma  \sum_{i=1}^n  w_i ~\! \frac{  e^{- \frac{(X_i-\ell)^2}{2 \xi^2}}(X_i-\ell)}{2 \xi^3}
~\!  \left[
1 - \erf \left( - \frac{   \frac{\gamma}{\sigma} \left( \frac{\ell+m}{2}  -  X_i \right) + \frac{\sigma}{ \gamma} \frac{m-\ell}{2} }{\sqrt{2} ~\! \xi} \right)
\right]\\
&&+ \gamma  \sum_{i=1}^n  w_i ~\! \frac{  e^{- \frac{(X_i-\ell)^2}{2 \xi^2}}}{ \sqrt{8\pi} ~\! \xi^2}  ~
\left(   \frac{\gamma}{ \sigma} - \frac{\sigma}{ \gamma} \right) 
 \exp \left[ - \frac{ \left(  \frac{ \gamma}{\sigma} \left( \frac{\ell+m}{2}  -  X_i \right) + \frac{\sigma}{ \gamma}  \frac{m-\ell}{2}  \right)^2}{ 2 ~\! \xi^2} \right], \\
\nonumber
\frac{\partial S_M}{\partial m}(\ell,m, r,t) &=& 
\gamma  \sum_{i=1}^n  w_i ~\! \frac{  e^{- \frac{(X_i-m)^2}{2 \xi^2}}(X_i-m)}{2 \xi^3}
~\!  
\erf \left( - \frac{   \frac{\gamma}{\sigma} \left( \frac{\ell+m}{2}  -  X_i \right) + \frac{\sigma}{ \gamma} \frac{m-\ell}{2} }{\sqrt{2} ~\! \xi} \right)\\
\nonumber
&&- 
\gamma  \sum_{i=1}^n  w_i ~\! \frac{  e^{- \frac{(X_i-m)^2}{2 \xi^2}}(X_i-m)}{2 \xi^3}
~\!  \erf \left(  -\frac{   \frac{ \gamma}{\sigma} \left(  \frac{m+r}{2} -X_i \right) - \frac{\sigma}{\gamma} \frac{r-m}{2}}{ \sqrt{2} ~\! \xi} \right)\\
\nonumber
&&+ \gamma  \sum_{i=1}^n  w_i ~\! \frac{  e^{- \frac{(X_i-m)^2}{2 \xi^2}}}{ \sqrt{8\pi} ~\! \xi^2}  ~
\left(   \frac{\gamma}{ \sigma} - \frac{\sigma}{ \gamma} \right) 
 \exp \left[ - \frac{ \left(  \frac{ \gamma}{\sigma} \left( \frac{\ell+m}{2}  -  X_i \right) + \frac{\sigma}{ \gamma}  \frac{m-\ell}{2}  \right)^2}{ 2 ~\! \xi^2} \right] \\
&&- \gamma  \sum_{i=1}^n  w_i ~\! \frac{  e^{- \frac{(X_i-m)^2}{2\xi^2}}}{ \sqrt{8\pi} ~\! \xi^2}  ~
\left(  \frac{\gamma}{ \sigma} - \frac{\sigma}{\gamma} \right) 
 \exp \left[ - \frac{ \left(  \frac{\gamma}{\sigma} \left(  \frac{m+r}{2} - X_i  \right) - \frac{\sigma}{\gamma} \frac{r-m}{2} \right)^2}{ 2 \xi^2} \right],\\
\nonumber
\frac{\partial S_R}{\partial r}(\ell,m,r,t) &=& 
 \gamma  \sum_{i=1}^n  w_i ~\! \frac{  e^{- \frac{(X_i-r)^2}{2\xi^2}}(X_i-r)}{ 2\xi^3}
~\!  \left[
1 + \erf \left(  -\frac{   \frac{ \gamma}{\sigma} \left(  \frac{m+r}{2} -X_i \right) - \frac{\sigma}{\gamma} \frac{r-m}{2}}{ \sqrt{2} ~\! \xi} \right)
\right]\\
&&- \gamma  \sum_{i=1}^n  w_i ~\! \frac{  e^{- \frac{(X_i-r)^2}{2\xi^2}}}{ \sqrt{8\pi} ~\! \xi^2}  ~
\left(  \frac{\gamma}{ \sigma} - \frac{\sigma}{\gamma} \right) 
 \exp \left[ - \frac{ \left(  \frac{\gamma}{\sigma} \left(  \frac{m+r}{2} - X_i  \right) - \frac{\sigma}{\gamma} \frac{r-m}{2} \right)^2}{ 2 \xi^2} \right].\label{eq:3C-dS_R_evaluated}
\end{eqnarray}}

\begin{figure}[h]
    \centering
    \includegraphics[width=0.6\linewidth]{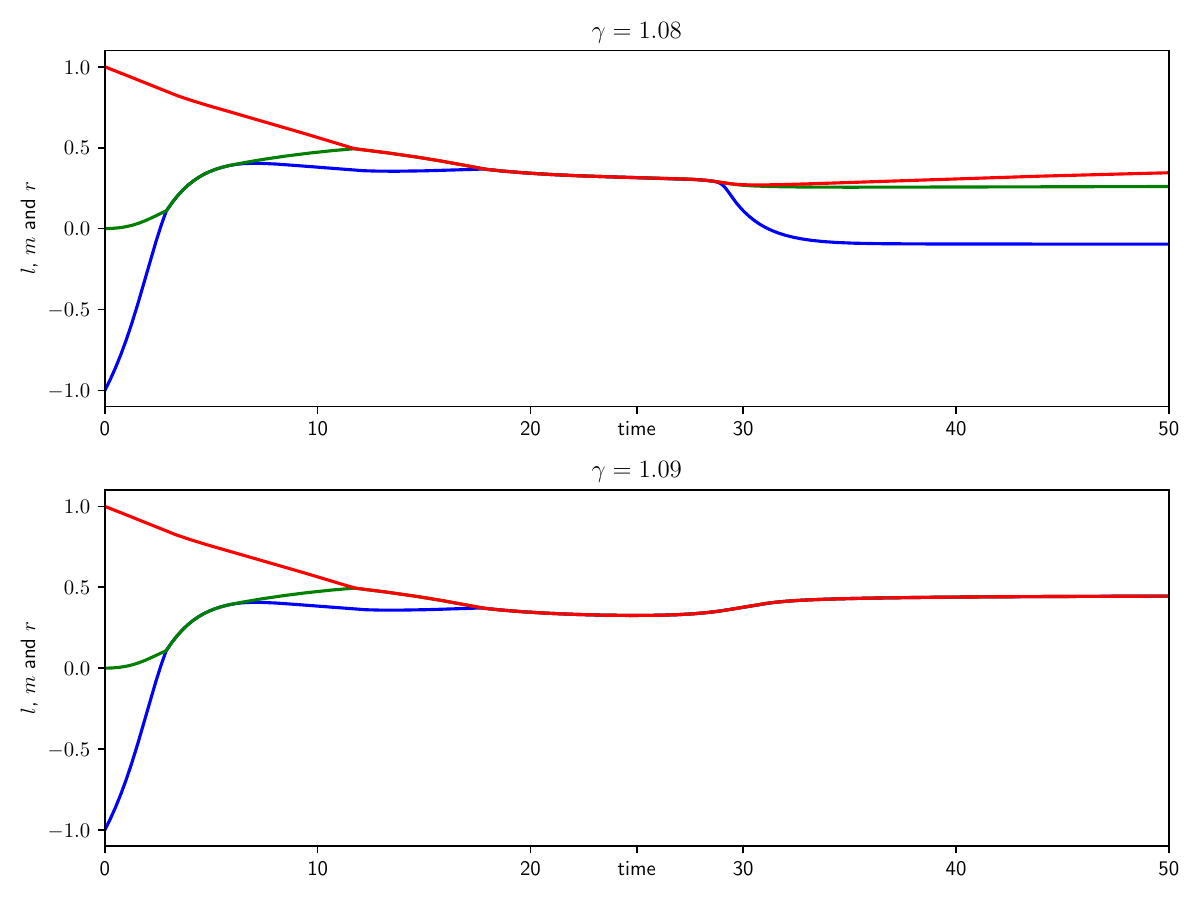}
    \caption{\textbf{Varying voter loyalty in 3-candidate setting.} Positions taken by left, middle, and right candidates as a function of time, with $\nu=0.25$, $\alpha=1$, $\kappa=\beta=0.2$, and two different values of $\gamma$.}
    \label{fig:three-candidate}
\end{figure}

In Fig.~\ref{fig:three-candidate}, we fixed voter open-mindedness $\nu = 0.25$, candidate opportunism $\alpha=1$ and $\kappa=\beta=0.2$, and varied voter loyalty $\gamma$. We observe a more complex behavior with the three candidates compared to the two-candidate situation. In both examples, we see the middle and left candidates coming together for a short time before diverging again, indicating a temporary coalition that arises under specific conditions of voter loyalty. This suggests that multilateral interactions between candidates can lead to fluctuating alliances, influenced by voters' varying degrees of loyalty and preferences. Next, middle and right candidates coalesce, and after a short time their position converges together with left candidate as well.

The difference depending on varying  voter loyalty coupled with increasing polarization is that in the larger loyalty $\gamma$ case the candidates remain together, while smaller gamma first causes left candidate to decouple from the coalition, and then after a short time the right candidate also detaches, leading to a more fragmented political landscape. 


\section{Discussion and extensions.} This paper presents a first study of candidate-voter dynamics, and the conclusion is remarkable: The optimal candidate 
position can depend discontinuously on the political environment. Countless variations would be of interest as well. For instance: 

\vskip 10pt
\begin{enumerate}
\item In this study, we only considered the influence of the electorate on the candidates, not the influence of charismatic candidates on the electorate. This is an obviously important effect in real politics, and we will add it to the model in future work. 
\item 
In \cite{Borgers_et_al_candidate_dynamics}, where candidate dynamics under static electorate was studied, the bifurcation diagrams were used to better illustrate the discontinuity in the behavior. The same effect is observed under dynamic electorate, which we illustrated in Fig.~\ref{fig:FINAL_CANDIDATE_POSITION} showing discontinuous dependence of final candidate position on voter loyalty. 
It is important to note, that the presence of this discontinuity rests on the fact that the voter population did not reach consensus due to the finite support of the voter interaction function \cite{Motsch_Tadmor}. 
Future work will include a deeper exploration of these effects.
\item Voters hold opinions about more than one issue, and consequently the opinion space ought to be higher-dimensional. 
\item In this study we also considered the more complex dynamic interaction between three candidates and effect of voter dynamics and loyalty. We observed the possible creation and fragmentation of coalitions. Future research should explore the long-term effects of these dynamics on party strategies and voter alignment. Understanding how candidates adapt their narratives and policies in response to shifting voter loyalties will be crucial in understanding electoral results in increasingly polarized environments. In addition, studying the role of social media and other digital platforms will provide valuable insights into the evolution of political communication. The intersection between technology and politics is rapidly changing the way candidates engage with the electorate, potentially amplifying not only their reach but also the intensity of political polarization. 
\item There is a limit to voters' willingness to adopt other people's opinions --- they have ``inherent" views from which they may not be willing to 
stray arbitrarily far. Such effects can easily be incorporated into our model. 
\item Voter views may be subject to random fluctuations, because of accidental events in their lives. 
This could be counteracted by adding {\em diffusion} (spontaneous random small changes in opinions) 
in the model, as some  authors  have proposed (see for instance \cite{Ben_Naim_2005,Goddard_2022}). This would
 raise the question how to incorporate diffusion in the particle method. One possibility 
is to include the convolutions with Gaussian kernels, which we only perform at the final time for plotting purposes, in each time step (or in
any case more often than once) and re-distribute the $X_i$ accordingly. 

\item 
Our model starts with the original Hegselmann-Krause model \cite{hegsel_krause_2002}.
In a later paper \cite{Hegselmann_Krause_2005}, Hegselmann and Krause suggested that individuals might respond not to 
the arithmetic average (or, in our modification of the model, weighted arithmetic average) of opinions in their vicinity, but to 
a different kind of average --- geometric averages for instance. We have not yet thought about what would happen if
we followed this interesting suggestion in our model.
\item We acknowledge that real elections are complex events with many elements that the model does not entirely capture; nonetheless, it remains insightful, and we intend to expand the model in the future. Some ideas for enhancing the model include incorporating more dimensions and performing parameter fitting of a real election to assess how well the model replicates it. For future work, we plan to use data from the Pew Research Center on the polarization of the American public between 1994 and 2014 \cite{pew_research_2014} to conduct a parameter fitting of the model.

\end{enumerate}

\vskip 10pt
{\bf Model Documentation.} The code used to implement the simulations and analyze the outcomes of the model is publicly available on \href{https://github.com/akirshtein/Candidate-Voter-Dynamics}{GitHub}, allowing for transparency and reproducibility of our results: \url{https://github.com/akirshtein/Candidate-Voter-Dynamics}.
\bigskip

\noindent
{\bf Acknowledgment.} This work was supported in part by a seed grant from the Data Intensive Studies Center at Tufts University. 

\newpage
\section{Appendix}
\subsection{Derivation of formulas \eqref{eq:S_L_evaluated}-\eqref{eq:dS_R_evaluated}}
\underline{Derivation of eq.\ (15) in the main text:} 

\begin{eqnarray} 
\nonumber
S_L(\ell,r,t) &=& \int_{-\infty}^{(\ell+r)/2} f(x,t) e^{-(x-\ell)^2/(2 \gamma^2)} ~\! dx \\
\nonumber
&=&  \int_{-\infty}^{(\ell+r)/2} \left( \sum_{i=1}^n w_i \frac{e^{-(x-X_i)^2/(2 \sigma^2)}}{\sqrt{2 \pi \sigma^2}} \right) e^{-(x-\ell)^2/(2 \gamma^2)} ~\! dx \\
\label{die_dritte} 
&=& \sum_{i=1}^n \frac{w_i}{\sqrt{2 \pi \sigma^2}} 
\int_{-\infty}^{(\ell+r)/2}{e^{-(x-X_i)^2/(2 \sigma^2)}}~\! e^{-(x-\ell)^2/(2 \gamma^2)} ~\! dx
\end{eqnarray}
where we have written $X_i$ instead of $X_i(t)$ for brevity. We denote the integral in \eqref{die_dritte} by 
$I_i$:
\begin{equation}
\label{eq:def_I_i}
I_i  = 
\int_{-\infty}^{(\ell+r)/2}{e^{-(x-X_i)^2/(2 \sigma^2)}}~\! e^{-(x-\ell)^2/(2 \gamma^2)} ~\! dx.
\end{equation}
We will compute $I_i$ now. We 
have 
\begin{equation}
\label{I_i_simplified_is_what} 
I_i = \int_{-\infty}^A e^{-a x^2 + bx - c} ~\! dx
\end{equation}
with
\begin{eqnarray} 
\label{definition_von_A}
A &=& \frac{\ell+r}{2}, \\
\label{definition_von_a} 
a &=& \frac{1}{2 \sigma^2} + \frac{1}{2 \gamma^2}, \\
b &=& \frac{X_i}{\sigma^2} + \frac{\ell}{\gamma^2}, \\
\label{definition_von_c}
c &=& \frac{X_i^2}{2 \sigma^2} + \frac{\ell^2}{2 \gamma^2}.
\end{eqnarray}
Let's compute the integral in \eqref{I_i_simplified_is_what} then: 
\begin{eqnarray}
\label{completed_the_square0}
I_i = 
\int_{-\infty}^A e^{-a x^2 + bx - c} ~\! dx = 
\int_{-\infty}^A e^{-\left(\sqrt{a}~\!  x - \frac{b}{2 \sqrt{a}} \right)^2} ~\! dx ~ e^{\frac{b^2}{4a} - c}. 
\end{eqnarray}
Substitute 
$$
s = \sqrt{a}~\!  x - \frac{b}{2 \sqrt{a}} 
$$
in \eqref{completed_the_square0}: 
\begin{eqnarray}
\nonumber
I_i &=& \int_{-\infty}^{\sqrt{a} ~\! A - \frac{b}{2 \sqrt{a}}} e^{-s^2} \frac{ds}{\sqrt{a}} ~\! e^{\frac{b^2}{4a} - c} \\
\nonumber
&=& \frac{e^{\frac{b^2}{4a}-c}}{\sqrt{a}} \int_{-\infty}^{\sqrt{a} ~\! A - \frac{b}{2 \sqrt{a}}} e^{-s^2} ~\! ds \\
\nonumber
&=& \frac{e^{\frac{b^2}{4a}-c}}{\sqrt{a}} \left( \frac{\sqrt{\pi}}{2} + \frac{ \sqrt{\pi}}{2} \erf \left( \sqrt{a} ~\! A - \frac{b}{2 \sqrt{a}} \right) \right) \\
\label{eq:I_i_almost_there} 
&=&  \frac{\sqrt{\pi} ~e^{\frac{b^2}{4a}-c}}{2\sqrt{a}}  \left( 1 + \erf \left( \sqrt{a}~\! A - \frac{b}{2 \sqrt{a}} \right) \right).
\end{eqnarray}
We first evaluate the factor 
$$
\frac{\sqrt{\pi}}{2 \sqrt{a}}
$$
appearing in \eqref{eq:I_i_almost_there}. Using the definition of $a$, equation \eqref{definition_von_a}, we have
\begin{eqnarray}
\nonumber
\frac{\sqrt{\pi}}{2 \sqrt{a}} &=& 
\frac{\sqrt{\pi}}{2 \sqrt{ \frac{1}{2 \sigma^2} + \frac{1}{2 \gamma^2}}} \\ 
\nonumber &=& \frac{\sqrt{\pi}}{ \sqrt{ \frac{2}{ \sigma^2} + \frac{2}{ \gamma^2}}} \\ 
\nonumber &=& \gamma~\!  \sigma~\!  \sqrt{\frac{\pi}{2}} ~ \frac{1}{\sqrt{\sigma^2 + \gamma^2}}.
\end{eqnarray}
Writing 
$$
\xi = \sqrt{\sigma^2+\gamma^2}, 
$$
we find: 
\begin{equation}
\label{first_fraction} 
\frac{\sqrt{\pi}}{2 \sqrt{a}}  =  \frac{ \gamma~\!  \sigma}{\xi} ~\!  \sqrt{\frac{\pi}{2}} .
\end{equation}

Next we evaluate the expression $\frac{b^2}{4a}-c$ appearing in the exponent in \eqref{eq:I_i_almost_there}.
Using the definitions of $a$, $b$, and $c$, we have 
\begin{eqnarray}
\nonumber
\frac{b^2}{4a}-c &=&  \frac{ \left(  \frac{X_i}{\sigma^2} + \frac{\ell}{\gamma^2} \right)^2}{4 \left( \frac{1}{2 \sigma^2} + \frac{1}{2 \gamma^2} \right) }- \left(  \frac{X_i^2}{2 \sigma^2} + \frac{\ell^2}{2 \gamma^2} \right) \\
&=& \frac{ \left(  \frac{X_i}{2\sigma^2} + \frac{\ell}{2\gamma^2} \right)^2}{\left( \frac{1}{2 \sigma^2} + \frac{1}{2 \gamma^2} \right) }- \left(  \frac{X_i^2}{2 \sigma^2} + \frac{\ell^2}{2 \gamma^2} \right). 
\label{hier_ist_der_exponent} 
\end{eqnarray}
We write 
$$
\alpha = \frac{1}{2 \sigma^2} ~~~\mbox{and} ~~~ \beta = \frac{1}{2 \gamma^2}.
$$
With this notation, \eqref{hier_ist_der_exponent} becomes 

\begin{align}
&
\frac{(\alpha X_i + \beta \ell)^2}{\alpha+\beta} - \alpha X_i^2 - \beta \ell^2 \nonumber\\
&= \frac{\alpha^2 X_i^2 + 2 \alpha \beta X_i \ell + \beta^2 \ell^2 - (\alpha+\beta) (\alpha X_i^2 + \beta \ell^2)}{\alpha + \beta}  \nonumber\\
&= \frac{-\alpha \beta X_i^2 - \alpha \beta \ell^2 + 2 \alpha \beta X_i \ell}{\alpha + \beta} \nonumber\\
&= - \frac{\alpha \beta}{\alpha + \beta} ~\! (X_i - \ell)^2 
= - \frac{1}{ \frac{1}{\alpha} + \frac{1}{\beta}} (X_i-\ell)^2 \nonumber\\
&= - \frac{1}{2 (\sigma^2+\gamma^2)} (X_i - \ell)^2 
= - \frac{(X_i-\ell)^2}{2 \xi^2}.\label{must_be_same} 
\end{align}
Finally we evaluate the expression $\sqrt{a} ~\! A - \frac{b}{2 \sqrt{a}}$ appearing in \eqref{eq:I_i_almost_there}:

\begin{eqnarray}
\nonumber
\sqrt{a} ~\! A - \frac{b}{2 \sqrt{a}} &=& \sqrt{\frac{1}{2 \sigma^2} + \frac{1}{2 \gamma^2}} ~\!  \frac{\ell+r}{2} - 
\frac{\frac{X_i}{\sigma^2} + \frac{\ell}{\gamma^2}}{2 \sqrt{\frac{1}{2 \sigma^2} + \frac{1}{2 \gamma^2}}} \\
\nonumber
&=& \frac{1}{\sigma ~\! \gamma}  \sqrt{\frac{\sigma^2+\gamma^2}{2}}~\! \frac{\ell+r}{2}  - \frac{\sigma \gamma}{\sqrt{2}} 
\frac{\frac{X_i}{\sigma^2} + \frac{\ell}{\gamma^2}}{ \sqrt{\sigma^2+\gamma^2}} \\
&=& \frac{1}{\sigma ~\! \gamma}  \sqrt{\frac{\sigma^2+\gamma^2}{2}}~\! \frac{\ell+r}{2}  - \frac{\sigma \gamma}{\sqrt{2}} 
\frac{\frac{X_i}{\sigma^2} + \frac{\ell+r}{2\gamma^2}- \frac{r-\ell}{2 \gamma^2}}{ \sqrt{\sigma^2+\gamma^2}} 
\label{awful0} 
\end{eqnarray}
In \eqref{awful0}, $X_i$ is multiplied by 
$$
- \frac{\gamma}{\sigma} ~\! \frac{1}{\sqrt{2} ~\! \xi}.
$$
Further, $\frac{\ell+r}{2}$ is multiplied by 
$$
\frac{\xi}{\sqrt{2} ~\! \sigma ~\! \gamma} - \frac{\sigma}{ \sqrt{2} \gamma~\! \xi}  = 
\frac{\xi^2/\gamma - \sigma^2/\gamma}{\sqrt{2} ~\! \sigma~\! \xi}  = \frac{\gamma}{\sigma} ~\! \frac{1}{\sqrt{2} ~\! \xi}.
$$
And $\frac{r-\ell}{2}$ is multiplied by 

$$
\frac{\sigma \gamma}{\sqrt{2}} ~\! \frac{1}{\xi \gamma^2} =
\frac{\sigma} {\gamma}~\! \frac{1}{\sqrt{2} ~\! \xi}.
$$
So altogether \eqref{awful0} becomes: 
\begin{equation}
\label{awful_again} 
\sqrt{a} ~\! A  - \frac{b}{2 \sqrt{a}} = \frac{\gamma}{\sigma}  \frac{ \frac{\ell+r}{2} - X_i}{\sqrt{2}~\!  \xi } + \frac{\sigma}{\gamma} \frac{\frac{r-\ell}{2}}{\sqrt{2} ~\! \xi}.
\end{equation}
So \eqref{eq:I_i_almost_there} becomes
\begin{equation}
\label{eq:I_i_have_it} 
I_i = \frac{ \gamma~\!  \sigma}{\xi} ~\!  \sqrt{\frac{\pi}{2}} e^{ - \frac{(X_i-\ell)^2}{2 \xi^2}}
\left( 1 + \erf \left( \frac{\gamma}{\sigma}  \frac{ \frac{\ell+r}{2} - X_i}{\sqrt{2}~\!  \xi } + \frac{\sigma}{\gamma} \frac{\frac{r-\ell}{2}}{\sqrt{2} ~\! \xi} \right) \right).
\end{equation}
Combining this with \eqref{die_dritte}, we find: 
{\footnotesize\begin{eqnarray}
\nonumber
S_L(\ell,r,t) &=& \sum_{i=1}^n \frac{w_i}{\sqrt{2 \pi \sigma^2}} ~\! 
\frac{ \gamma~\!  \sigma}{\xi} ~\!  \sqrt{\frac{\pi}{2}} e^{ - \frac{(X_i-\ell)^2}{2 \xi^2}}
\left( 1 + \erf \left( \frac{\gamma}{\sigma}  \frac{ \frac{\ell+r}{2} - X_i}{\sqrt{2}~\!  \xi } + \frac{\sigma}{\gamma} \frac{\frac{r-\ell}{2}}{\sqrt{2} ~\! \xi} \right) \right) \\
\nonumber
&=&  \frac{\gamma}{2 \xi} \sum_{i=1}^n w_i ~\!   e^{ - \frac{(X_i-\ell)^2}{2 \xi^2}}
\left( 1 + \erf \left( \frac{\gamma}{\sigma}  \frac{ \frac{\ell+r}{2} - X_i}{\sqrt{2}~\!  \xi } + \frac{\sigma}{\gamma} \frac{\frac{r-\ell}{2}}{\sqrt{2} ~\! \xi} \right) \right)  \\
\label{eq:SL}
&=&  \frac{\gamma}{2 \xi} \sum_{i=1}^n w_i ~\!   e^{ - \frac{(X_i-\ell)^2}{2 \xi^2}}
\left( 1 - \erf \left( \frac{\gamma}{\sigma}  \frac{X_i- \frac{\ell+r}{2} }{\sqrt{2}~\!  \xi } - \frac{\sigma}{\gamma} \frac{\frac{r-\ell}{2}}{\sqrt{2} ~\! \xi} \right) \right)  
\end{eqnarray}}
\newpage
\noindent
\underline{Derivation of eq.\ (16) in the main text:} 

\begin{eqnarray} 
\nonumber
S_R(\ell,r,t) &=& \int_{(\ell+r)/2}^\infty f(x,t) e^{-(x-r)^2/(2 \gamma^2)} ~\! dx \\
\nonumber
&=&  \int_{(\ell+r)/2}^\infty \left( \sum_{i=1}^n w_i \frac{e^{-(x-X_i)^2/(2 \sigma^2)}}{\sqrt{2 \pi \sigma^2}} \right) e^{-(x-r)^2/(2 \gamma^2)} ~\! dx \\
\label{die_dritte_wieder} 
&=& \sum_{i=1}^n \frac{w_i}{\sqrt{2 \pi \sigma^2}} 
 \int_{(\ell+r)/2}^\infty {e^{-(x-X_i)^2/(2 \sigma^2)}}~\! e^{-(x-r)^2/(2 \gamma^2)} ~\! dx
\end{eqnarray}
where we have again written $X_i$ instead of $X_i(t)$ for brevity. We now denote the integral in \eqref{die_dritte_wieder} by 
$J_i$:
\begin{equation}
\label{eq:def_I_i_again} 
J_i = 
 \int_{(\ell+r)/2}^\infty {e^{-(x-X_i)^2/(2 \sigma^2)}}~\! e^{-(x-r)^2/(2 \gamma^2)} ~\! dx.
\end{equation} 
Substitute $s=-x$, then write $x$ instead of $s$ again: 
$$
J_i = 
\int_{-\infty}^{-(\ell+r)/2} e^{-(x+X_i)^2/(2 \sigma^2)} e^{- (x+r)^2/(2 \gamma^2)} ~\! dx. 
$$
This is the same as \eqref{eq:def_I_i}, except that $\ell$ has been replaced by $-r$, 
$r$ has been replaced by $-\ell$, and $X_i$ has been replaced by $-X_i$. 
Making these substitutions in \eqref{eq:I_i_have_it}, we obtain
\begin{eqnarray}
\label{eq:I_i_have_it_2} 
J_i &=&  
 \frac{ \gamma~\!  \sigma}{\xi} ~\!  \sqrt{\frac{\pi}{2}} e^{ - \frac{(X_i-r)^2}{2 \xi^2}}
\left( 1 + \erf \left( - \frac{\gamma}{\sigma}  \frac{ \frac{\ell+r}{2} - X_i}{\sqrt{2}~\!  \xi } + \frac{\sigma}{\gamma} \frac{\frac{r-\ell}{2}}{\sqrt{2} ~\! \xi} \right) \right) 
\end{eqnarray}

Combining this with \eqref{die_dritte_wieder}, we find: 
{\footnotesize\begin{eqnarray}
\nonumber
S_R(\ell,r,t) &=& \sum_{i=1}^n \frac{w_i}{\sqrt{2 \pi \sigma^2}} ~\! 
\frac{ \gamma~\!  \sigma}{\xi} ~\!  \sqrt{\frac{\pi}{2}} e^{ - \frac{(X_i-r)^2}{2 \xi^2}}
\left( 1 + \erf \left(  \frac{\gamma}{\sigma}  \frac{ X_i - \frac{\ell+r}{2} }{\sqrt{2}~\!  \xi } + \frac{\sigma}{\gamma} \frac{\frac{r-\ell}{2}}{\sqrt{2} ~\! \xi} \right) \right)\\
\label{eq:SR} 
&=&  \frac{\gamma}{2 \xi} \sum_{i=1}^n w_i ~\!   e^{ - \frac{(X_i-r)^2}{2 \xi^2}}
\left( 1 + \erf \left(  \frac{\gamma}{\sigma}  \frac{ X_i - \frac{\ell+r}{2} }{\sqrt{2}~\!  \xi } - \frac{\sigma}{\gamma} \frac{\frac{\ell-r}{2}}{\sqrt{2} ~\! \xi} \right) \right) 
\end{eqnarray}}
Notice that \eqref{eq:SR} is obtained from \eqref{eq:SL} by swapping $\ell$ with $r$, and replacing $-\erf$ by $+\erf$.

\newpage
\noindent
\underline{Derivation of eq.\ (17) in the main text:} 

\vskip 10pt
Using eq.\ \eqref{eq:SL}, we now compute: 
\[
\frac{\partial S_L}{\partial \ell} = \frac{\partial }{\partial \ell}  \left( 
\frac{\gamma}{2 \xi} \sum_{i=1}^n w_i ~\!   e^{ - \frac{(X_i-\ell)^2}{2 \xi^2}}
\left( 1 - \erf \left( \frac{\gamma}{\sigma}  \frac{X_i- \frac{\ell+r}{2} }{\sqrt{2}~\!  \xi } - \frac{\sigma}{\gamma} \frac{\frac{r-\ell}{2}}{\sqrt{2} ~\! \xi} \right) \right)  \right) 
\]
Omit the factor of $\frac{\gamma}{2 \xi}$, the summation, and the factor of $w_i$ for now: 
{\footnotesize\begin{align*}
&  \frac{\partial }{\partial \ell}  \left(   e^{ - \frac{(X_i-\ell)^2}{2 \xi^2}}
\left( 1 - \erf \left( \frac{\gamma}{\sigma}  \frac{X_i- \frac{\ell+r}{2} }{\sqrt{2}~\!  \xi } - \frac{\sigma}{\gamma} \frac{\frac{r-\ell}{2}}{\sqrt{2} ~\! \xi} \right) \right)  \right)  \\
&=  \frac{X_i-\ell}{\xi^2}~e^{ - \frac{(X_i-\ell)^2}{2 \xi^2}} ~\left( 1 - \erf \left( \frac{\gamma}{\sigma}  \frac{X_i- \frac{\ell+r}{2} }{\sqrt{2}~\!  \xi } - \frac{\sigma}{\gamma} \frac{\frac{r-\ell}{2}}{\sqrt{2} ~\! \xi} \right) \right) \\
& ~~+ \frac{2}{\sqrt{\pi}} ~ e^{ - \frac{(X_i-\ell)^2}{2 \xi^2}} ~\exp \left( - \left( \frac{\gamma}{\sigma}  \frac{X_i- \frac{\ell+r}{2} }{\sqrt{2}~\!  \xi } - \frac{\sigma}{\gamma} \frac{\frac{r-\ell}{2}}{\sqrt{2} ~\! \xi} \right)^2 \right) ~ \left(  \frac{\gamma}{2 \sqrt{2} ~\sigma ~ \xi}  - \frac{\sigma}{2 \sqrt{2} ~\gamma~  \xi} \right) \\ 
&=  \frac{X_i-\ell}{\xi^2}~e^{ - \frac{(X_i-\ell)^2}{2 \xi^2}} ~\left( 1 - \erf \left( \frac{\gamma}{\sigma}  \frac{X_i- \frac{\ell+r}{2} }{\sqrt{2}~\!  \xi } - \frac{\sigma}{\gamma} \frac{\frac{r-\ell}{2}}{\sqrt{2} ~\! \xi} \right) \right) \\
& ~~+ \frac{1}{\sqrt{2 \pi}~ \xi} ~ e^{ - \frac{(X_i-\ell)^2}{2 \xi^2}} ~\exp \left( - \left( \frac{\gamma}{\sigma}  \frac{X_i- \frac{\ell+r}{2} }{\sqrt{2}~\!  \xi } - \frac{\sigma}{\gamma} \frac{\frac{r-\ell}{2}}{\sqrt{2} ~\! \xi} \right)^2 \right) ~ \left( \frac{\gamma}{\sigma} - \frac{\sigma}{\gamma} \right) 
\end{align*} }
Bringing back the factor of $\frac{\gamma}{2 \xi}$, the summation, and the factor of $w_i$ now: 
{\scriptsize\begin{eqnarray}
\nonumber
\frac{\partial S_L}{\partial \ell} &=&  \gamma~ \sum_{i=1}^n  ~w_i~  \frac{e^{ - \frac{(X_i-\ell)^2}{2 \xi^2}} (X_i-\ell)}{2 \xi^3} 
\left[ 1 - \erf \left( - \frac{ \frac{\gamma}{\sigma}   \left(  \frac{\ell+r}{2}-X_i \right)   + \frac{\sigma}{\gamma} \frac{r-\ell}{2} }{\sqrt{2}~ \xi} \right) \right] \\
\label{eq:dell_SL_by_dell_ell} 
&~& ~~~~+  \gamma~ \sum_{i=1}^n  ~w_i~  \frac{e^{ - \frac{(X_i-\ell)^2}{2 \xi^2}} }{\sqrt{8 \pi}  ~ \xi^2} ~ \left( \frac{\gamma}{\sigma} - \frac{\sigma}{\gamma} \right) 
~\exp \left( - \frac{ \left( \frac{\gamma}{\sigma}   \left(  \frac{\ell+r}{2} - X_i  \right) + \frac{\sigma}{\gamma} \frac{r-\ell}{2} \right)^2}{2~ \xi^2} \right) 
\end{eqnarray} }

\noindent
\underline{Derivation of eq.\ (18) in the main text:} 

\vskip 10pt
We pointed out earlier that 
\eqref{eq:SR} is obtained from \eqref{eq:SL} by swapping $\ell$ with $r$, and replacing $-\erf$ by $+\erf$.
This, together with \eqref{eq:dell_SL_by_dell_ell}, implies 

{\scriptsize\begin{eqnarray}
\nonumber
\frac{\partial S_R}{\partial r} &=&  \gamma~ \sum_{i=1}^n  ~w_i~  \frac{e^{ - \frac{(X_i-r)^2}{2 \xi^2}} (X_i-r)}{2 \xi^3} 
\left[ 1 + \erf \left( - \frac{ \frac{\gamma}{\sigma}   \left(  \frac{\ell+r}{2}-X_i \right)   - \frac{\sigma}{\gamma} \frac{r-\ell}{2} }{\sqrt{2}~ \xi} \right) \right] \\
\label{eq:dell_SL_by_dell_ell} 
&~& ~~~~ - \gamma~ \sum_{i=1}^n  ~w_i~  \frac{e^{ - \frac{(X_i-r)^2}{2 \xi^2}} }{\sqrt{8 \pi}  ~ \xi^2} ~ \left( \frac{\gamma}{\sigma} - \frac{\sigma}{\gamma} \right) 
~\exp \left( - \frac{ \left( \frac{\gamma}{\sigma}   \left(  \frac{\ell+r}{2} - X_i  \right) - \frac{\sigma}{\gamma} \frac{r-\ell}{2} \right)^2}{2~ \xi^2} \right) 
\end{eqnarray}}

Deriving the formulas \eqref{eq:3C-S_L_evaluated}-\eqref{eq:3C-dS_R_evaluated} is analogous and therefore is omitted.


\small{

}

\end{document}